\documentclass[aps,prd,twocolumn,showpacs,preprintnumbers,amsmath,amssymb,nofootinbib]{revtex4-1}
\usepackage{amsmath,amsfonts,euscript,amssymb}
\usepackage{epsfig}
\DeclareSymbolFont{bfitletters}{OML}{cmm}{bx}{it}
\DeclareSymbolFont{bfitoperators}   {OT1}{cmr} {m}{n}
\DeclareMathSymbol{\bfitomega}{\mathord}{bfitletters}{"21}
\DeclareMathSymbol{\bfitrho}{\mathord}{bfitletters}{"1A}
\DeclareMathSymbol{\bfitLambda}{\mathord}{operators}{"03}
\DeclareMathSymbol{\lambda}{\mathord}{letters}{"15}
\DeclareMathSymbol{\bfitgamma}{\mathord}{bfitletters}{"0D}

\newcommand{\be}{\begin{equation}}
\newcommand{\ee}{\end{equation}}
\newcommand{\bea}{\begin{eqnarray}}
\newcommand{\eea}{\end{eqnarray}}
\begin{document}

\title{Extrapolation of Hubble Curves}

\author{\bf A. E. Pavlov}
\affiliation{Institute of Mechanics and Power Engineering, Russian State Agrarian University -- Moscow Timiryazev Agricultural Academy, Timiryazevskaya str. 49,
Moscow 127550, Russia\\
alexpavlov60@mail.ru}

\begin{abstract}
We present exact solutions to the Friedmann equation in coordinate time and conformal time in elliptic Legendre integrals. Both approaches describe the modern Hubble diagram with the same accuracy. Hubble curves are extrapolated for large redshift values.
\end{abstract}


\maketitle

\section{Introduction}

In recent years, two independent collaborations ``High Supernova'' and ``Supernova Cosmology Project'' obtained new unexpected data about cosmological evolution at very large distances -- hundreds and thousands megaparsecs expressed in redshift values $z=1\div 1.7$~\cite{RiessNobel}. Surprisingly, it was found that the decrease of brightness with distance happen noticeably faster than it was expected according to the standard cosmological model with the matter dominance. Supernovae are situated at distances further than it was predicted. Therefore, according to the standard cosmological model, in the last period, the cosmological expansion proceeds with acceleration. Dynamics, by unknown reasons passes from the deceleration stage to an acceleration one of expansion. Observable data testify that the Universe is filled mainly not with massive dust that can not provide accelerating expansion but with unspecified enigmatic substance of other nature -- ``dark energy''.

Theory of elementary particles fails to identify a theory for dark matter and dark energy. There are some free parameters that must be consistent with the experimental data.
The important purpose of testing a cosmological theory is vitiated~\cite{Ram}.
Recently, a large analysis of observational data on the expansion rate of the Universe at small $z$ has been carried out~\cite{Denzel, Soltis, RiessApJL}. It should be noted that the observational data for the Hubble constant differ.

Einstein's General Relativity was formulated in redundant variables. It is a theory with constraints, so it has been studied by methods of Hamiltonian dynamics. Friedmann equation that utilized for interpretations of the modern cosmological data is the Hamiltonian constraint by its meaning.
In conformal metric, the Universe is not expanding but the elementary particle's masses are depended on time~\cite{Narlikar}. Both approaches describe the modern Hubble diagram with the same accuracy~\cite{Zakharov:2010nf}. The theoretical curves interpolating the Hubble diagram are expressed in analytical form. The functions belong to the class of meromorphic Weierstrass functions~\cite{PavlovMIPh}.
Of interest is the extrapolation of the Hubble curves for large redshifts.

\section{Friedmann equation in classical cosmology}

In the standard cosmology the Friedmann equation is used for fitting {\rm SNe Ia} data.
The closed cosmological model is set by space interval
\be\label{interval}
\gamma_{ij}{\rm d}x^i{\rm d}x^j=a^2(t)\left({\rm d}\chi^2+\sin^2\chi ({\rm d}\theta^2+\sin^2\theta{\rm d}\phi^2)\right),
\ee
where $a(t)$ is the radius of the Universe as a function of the coordinate time $t$, $(\chi, \theta, \phi)$ are dimensionless angle variables. According to the Hamiltonian theory of gravity, the Friedmann equation is a Hamiltonian constraint~\cite{Pavlov2017}:
\be\label{Hamconstraint}
{\cal H}_\bot :=\sqrt\gamma\left(K_{ij}K^{ij}-K^2-R+16\pi G\rho\right).
\ee
Here, $\gamma$ is the determinant of the 3-sphere metric, $R=6/a^2$ is the Ricci scalar of the 3-sphere, $K_{ij}$ are components of the extrinsic curvature tensor
\be\label{Kij}
K_{ij}:=-\frac{1}{2}\dot\gamma_{ij},
\ee
the dot denotes the derivative with respect to the coordinate time $t$, $G$ is the Newton constant, $\rho$ is the matter density,
the trace of the extrinsic curvature tensor (\ref{Kij}) is
\be\label{traceK}
K:=\gamma^{ij}K_{ij}=-3\left(\frac{\dot{a}}{a}\right)\equiv 3H(t).
\ee
The quantity $H\equiv \dot{a}/a$ is the Hubble parameter defining the expansion rate of the Universe. Its present value $H_0$ calls the Hubble constant.
Hence, the Hamiltonian constraint (\ref{Hamconstraint}), after substitutions (\ref{Kij}), (\ref{traceK}) takes the well-known form of the Friedmann equation:
\begin{equation}\label{Friedmann}
\left(\frac{\dot{a}}{a}\right)^2=\frac{8\pi G}{3}\rho-\frac{1}{a^2},
\end{equation}

The energy continuity equation
\be\label{continuityclassical}
\dot{\rho}=-3(\rho+P)\left(\frac{\dot{a}}{a}\right),
\ee
with an equation of state $P=w\rho$, which connects the pressure $P$ and the energy density $\rho$, yields the dependence of the density of the scale factor. So,

$\bullet$
for interstellar dust $P=0: \rho\sim a^{-3};$

$\bullet$
for radiation $P=\rho/3: \rho\sim a^{-4};$

$\bullet$
for vacuum $P=-\rho: \rho\sim\Lambda;$

$\bullet$
for rigid state matter $P=\rho: \rho\sim a^{-6}.$

Introducing for the vacuum density, non-relativistic, relativistic, and rigid state matters the cosmological parameters $\Omega_{\rm \Lambda}$, $\Omega_{\rm M}$, $\Omega_{\rm rad},$ $\Omega_{\rm rigid},$ we present the energy density as the sum
\bea\label{rho}
&&\rho=\frac{3H_0^2}{8\pi G}\times\\
&\times&\left[
\Omega_{\rm\Lambda}+\Omega_{\rm M}\left(\frac{a_0}{a}\right)^3+
\Omega_{\rm rad}\left(\frac{a_0}{a}\right)^4+\Omega_{\rm rigid}\left(\frac{a_0}{a}\right)^6\right].\nonumber
\eea
The cosmological parameters are constrained
\begin{equation}\label{constraint}
\Omega_{\rm\Lambda}+\Omega_{\rm M}+\Omega_{\rm rad}+\Omega_{\rm rigid}+\Omega_{\rm curv}=1,
\end{equation}
where for curvature term is
\begin{equation}\label{curv}
\Omega_{\rm curv}\equiv -\frac{1}{a_0^2H_0^2}.
\end{equation}
Substituting (\ref{rho}), (\ref{constraint}), (\ref{curv}) into the right side of the Friedmann equation (\ref{Friedmann}), we obtain
\bea\label{Friedmannclass}
&&\frac{1}{x^2}\left(\frac{{\rm d}x}{{\rm d}t}\right)^2=\\
&=&H_0^2\left[\Omega_{\rm \Lambda}+\frac{\Omega_{\rm curv}}{x^2}+\frac{\Omega_{\rm M}}{x^3}+
\frac{\Omega_{\rm rad}}{x^4}+\frac{\Omega_{\rm rigid}}{x^6}\right].\nonumber
\eea
Here the variable $x$ is given as a ratio of a scale $a(t)$ to a modern value $a_0=1$:
\begin{equation}\label{x}
x\equiv\frac{a(t)}{a_0}=\frac{1}{1+z},
\end{equation}
$z$ is a redshift of spectral lines,
\be\label{Hubbleconstant}
H_0=h\cdot 10^5 m/s/Mpc,\quad h=0.72\pm 0.08
\ee
is the Hubble constant.
The data of modern astronomical observations are fitted using cosmological parameters ~\cite{RiessNobel}:
\begin{equation}
\Omega_\Lambda = 0.72,\quad \Omega_{\rm M}=0.28.
\end{equation}
The contributions of vacuum and baryonic matter prevail. The contributions of the remaining terms in the Friedmann equation (\ref{Friedmannclass}) are negligible.

Since the spacetime interval for a photon is zero, we have a relation between the intervals of time and space
$c{\rm d}t = -a(t){\rm d}r$.
Using the above notation (\ref{x}), we find the formula for the distance
\be\label{int}
-a_0 r=c\int\frac{{\rm d}t}{x}=c\int\frac{{\rm d}x}{x}\frac{1}{{\rm d}x/{\rm d}t}.
\ee
Substituting the Friedmann equation (\ref{Friedmannclass}) with significant cosmological parameters into the integrand (\ref{int}), we obtain the formula for calculating the distance to a luminous object
\be\label{auxiliary}
r(z)=\frac{c}{H_0\sqrt{\Omega_\Lambda}}\int\limits_{1/(1+z)}^1\frac{{\rm d}x}{\sqrt{x^4+ p^3 x}},
\ee
where we denoted the ratio of partial contributions as
$p\equiv \sqrt[3]{{\Omega_{\rm M}}/{\Omega_\Lambda}}.$
The integral
\be\label{auxiliarySC}
I=\int\frac{{\rm d}x}{\sqrt{x^4+ p^3 x}},
\ee
can be expressed with using elliptic integrals~\cite{Whittaker}. To do this, we represent the radical expression in the form of a product of second-order polynomials
\be\nonumber
x^4+ p^3 x =(x^2-p x+ p^2)(x^2+p x).
\ee
Next, we apply the linear fractional substitution $x\mapsto t$
\be\nonumber
x=\frac{(1+\sqrt{3} )t+(1-\sqrt{3})}{2(t+1)}p.
\ee
The differential will be replaced as follows
\be\nonumber
{\rm d}x=-\frac{\sqrt{3} p}{(t+1)^2}{\rm d}t,
\ee
and the integral (\ref{auxiliarySC}) becomes
\be
I=-\frac{2}{\sqrt{(2\sqrt{3}-3)}p}\int\frac{{\rm d}t}{\sqrt{(1+t^2)(b^2t^2-1)}},
\ee
where $b\equiv (2+\sqrt{3}).$
With the trigonometric substitution $t\mapsto \varphi$
\be\nonumber
bt=\sec\varphi\equiv\frac{1}{\cos\varphi}
\ee
the integrand is reduced to the standard form
\be\nonumber
\frac{{\rm d}t}{\sqrt{(1+t^2)(b^2t^2-1)}}=\sqrt{1-k^2}\frac{{\rm d}\varphi}{\sqrt{1-k^2\sin^2\varphi}},
\ee
where $k^2\equiv b^2/(1+b^2)$.
As a result, the definite integral (\ref{auxiliary}) is expressed in terms of an elliptic integral of the first kind according to Legendre:
\be\nonumber
F(\varphi, k):=\int\limits_0^\varphi\frac{{\rm d}\phi}{\sqrt{1-k^2\sin^2\phi}}
\ee
Hence, we obtain the formula desired
\be\label{LegandreI}
r (z)=\frac{2c}{\sqrt[4]{3}p\sqrt{\Omega_\Lambda}H_0}
\left[F(\varphi(z),k)-F(\varphi (0),k)\right].
\ee
Here $k^2$ is the modulus of the elliptic integral of the first kind according to Legendre
\be\nonumber
k\equiv\frac{2+\sqrt{3}}{\sqrt{1+(2+\sqrt{3})^2}},
\ee
and the angle $\varphi(z)$ is defined by the formula:
\be\nonumber
\varphi (z)={\arccos} \left(
\frac{(\sqrt{3}+1)p(1+z)-2}{(\sqrt{3}+2)
[(\sqrt{3}-1)p(1+z)+2]}
\right).
\ee

In standard cosmology, the luminosity distance $d_L (z)_{SC}$ is related to the coordinate distance $r$:
\begin{equation}\label{dLs}
d_L (z)_{SC}=(1+z)a_0 r(z).
\end{equation}
Modern observational cosmology is based on the Hubble diagram. Effective stellar magnitude -- redshift relationship
\begin{equation}\label{mMcl}
m(z)-M=5{\rm lg} [d_L(z)_{SC}]+{\cal M},
\end{equation}
is used to test cosmological theories ($d_L$ in megaparsecs) ~\cite{RiessNobel}. Here $m(z)$ is the apparent magnitude, $M$ is its absolute magnitude, and ${\cal M}=25$ is a constant.

\section{Conformal General Relativity}

The Einstein equations are not invariant under conformal transformations. This raises the questions, whether it is suitable to work with such equations globally on cosmological distances and times~\cite{Fock}. Conformal transformations shrink or  stretch the spacetime intervals between two points described by the same coordinate system
\be\nonumber
{\rm d}s^2=\Omega^2{\rm d}\bar{s}^2
\ee
on the conformal to each other manifolds ${\cal M}$ and ${\cal\bar{M}}$.
Under the conformal transformation the components of the spacetime metric and corresponding metric determinant transform as
\be\label{confmetricmn}
g_{\mu\nu}=\Omega^2\bar{g}_{\mu\nu}, \qquad \sqrt{-g}=\Omega^4\sqrt{-\bar{g}}.
\ee
The Ricci scalar of the spacetime transforms according to
\be\label{confRiccimn}
R=\Omega^{-2}\left(\bar{R}-6\frac{\bar\Box\Omega}{\Omega}\right),
\ee
where the d'Alembertian operator ${\bar\Box}$ is taken with respect to the conformally rescaled metric. In order to understand the problem of conformal invariance let us start our discussion with the Hilbert action~\cite{ProblemsGauge}
\be\label{Hilbert}
S_{\rm H}=\frac{1}{2}\int\,{\rm d}^4x\sqrt{-g}\frac{1}{6}R,
\ee
where one puts
\be\nonumber
\kappa^2\equiv 8\pi G=6.
\ee
The conformal transformation (\ref{confmetricmn}) together with (\ref{confRiccimn}) yields the Hilbert action transformed
\be\label{confHilbert}
\bar{S}_{\rm H}=\frac{1}{2}\int\,
{\rm d}^4x\sqrt{-\bar{g}}\Omega^{2}\left(\frac{1}{6}\bar{R}-\frac{\bar\Box\Omega}{\Omega}\right).
\ee
The action (\ref{confHilbert}) is not conformal invariant. The additional term in (\ref{confHilbert}) looks as a kinetic term of an arbitrary scalar field. The idea is to couple a massless scalar field $\Phi$  in order to absorb the conformal factor
\be\label{HilbertPhi}
S_{\rm H\Phi}=\frac{1}{2}\int\,{\rm d}^4x\sqrt{-g}\left(\frac{1}{6}\right)R\Phi^2.
\ee
After the conformal transformation one yields
\be\label{HilbertPhiconform}
\bar{S}_{\rm H\Phi}=\frac{1}{2}\int\,{\rm d}^4x\sqrt{-\bar{g}}
\left(\frac{1}{6}\bar{R}\bar{\Phi}^2-\frac{\bar\Box\Omega}{\Omega}\bar{\Phi}^2\right),
\ee
where we take into account the scalar field rescaled
\be\label{Phirescaled}
\Phi=\Omega^{-1}\bar\Phi,
\ee
so the conformal factor in the action (\ref{HilbertPhiconform}) is absorbed.

Then, let us consider the action for a massless scalar field
\be\label{SPhi}
S_{\Phi}=-\frac{1}{2}\int\, {\rm d}^4x\sqrt{-g}\Phi\Box\Phi.
\ee
After the conformal transformation (\ref{confmetric}) is applied, with taking into account the field transformation (\ref{Phirescaled}), one yields the transformed field action
\be\label{SPhitransformed}
\bar{S}_{\Phi}=-\frac{1}{2}\int\, {\rm d}^4x\sqrt{-\bar{g}}\left(\bar{\Phi}\bar\Box\bar{\Phi}-
\frac{\bar\Box\Omega}{\Omega}\bar{\Phi}^2\right),
\ee
where we have used the relation
\be\nonumber
\Box\Phi=\frac{1}{\Omega^3}\left(\bar\Box\bar\Phi-\frac{\bar\Box\Omega}{\Omega}\bar{\Phi}^2\right).
\ee
As we see the scalar field action (\ref{SPhitransformed}) is not conformal invariant.

Now, if we add the action of the massless scalar field (\ref{SPhitransformed}) with its action coupled with gravitation field (\ref{HilbertPhiconform}), we get the conformally invariant action
\be\label{total}
S=\frac{1}{2}\int\, {\rm d}^4x\sqrt{-{g}}\Phi\left(\frac{1}{6}R\Phi-\Box\Phi\right).
\ee
The action (\ref{total}) can be presented in a general form including the massive scalar field and after discarding the boundary term
\be\label{totalaction}
S=\frac{1}{2}\int\, {\rm d}^4x\sqrt{-{g}}\left(\frac{1}{6}R\Phi^2+
g^{\mu\nu}\nabla_\mu\Phi\nabla_\nu\Phi-m^2\Phi^2\right).
\ee
Here we used the expression for the covariant d'Alembertian
\be\nonumber
\Box\Phi\equiv\frac{1}{\sqrt{-g}}\frac{\partial}{\partial x^\mu}
\left(\sqrt{-g}g^{\mu\nu}
\frac{\partial}{\partial x^\nu}\Phi\right).
\ee
The action with the mass term (\ref{totalaction}) is conformally invariant if we take into account the mass scaling
\be\nonumber
m=\Omega^{-1}\bar{m}.
\ee
With the particular $1/6$ coefficient of $R$ (\ref{totalaction}), this action is equivalent to that of General Relativity~\cite{Deser}.
Dilaton theories are successfully developed (see, {\it e.g.}~\cite{ProblemsGauge,Deser,Feza}).

In this paper we consider the spatial conformal transformations~\cite{Pavlov2017}.
Under the conformal transformation the components of the space metric and corresponding metric determinant transform as
\be\label{confmetric}
\gamma_{ij}=\Psi^4\bar{\gamma}_{ij}, \qquad \sqrt{\gamma}=\Psi^6\sqrt{\bar{\gamma}}.
\ee
The components of extrinsic curvature transform as
\be\label{extrinsic}
K_{ij}=\Psi^{-2}A_{ij}+\frac{1}{3}\Psi^4\bar\gamma_{ij}K,
\ee
where $A_{ij}$ are components of the traceless part of the tensor $K_{ij}$, $K$ is its trace.

The definitions (\ref{confmetric}), (\ref{extrinsic}) are redundant. These definitions are invariant under the conformal transformation~\cite{Brown}
\bea
\bar\gamma_{ij}&\longrightarrow&\xi^4\bar\gamma_{ij},\nonumber\\
\Psi&\longrightarrow&\xi^{-1}\Psi,\nonumber\\
A_{ij}&\longrightarrow&\xi^{-2}A_{ij},\nonumber\\
K&\longrightarrow& K,\nonumber
\eea
for any field $\xi$. The determinant of the metric is unspecified. The metric tensor represents a conformal equivalent class of metrics~\cite{Brown}.

The Ricci scalar of the space transforms according to
\be\label{confRicci}
R=\Psi^{-4}\left(\bar{R}-8\frac{\bar\triangle\Psi}{\Psi}\right),
\ee
where the Laplacian ${\bar\triangle}$ is taken with respect to the conformally rescaled metric.
The conformal Hamiltonian constraint
\bea
&&{\cal H}_\bot=\nonumber\\
&&=\frac{1}{2}\bar\gamma A^{ij}A_{ij}-\frac{1}{3}\bar\gamma\Psi^{12}K^2-\frac{1}{2}\bar\gamma\Psi^8\bar{R}+4\bar\gamma\Psi^7\bar\triangle\Psi
\nonumber\\
&&+\bar\gamma\Psi^{12}\rho=0\label{confconstraint}
\eea
is invariant under the conformal transformation (\ref{confmetric}).

\section{Conformal Friedmann equation}

For studying dynamics of gravitational field, the transition to the conformal metric with components $\tilde\gamma_{ij}$ is implemented~(\ref{confmetric}).
The factor $\Psi$ is defined as a power of the ratio of the spatial metric (\ref{interval}) determinant $\gamma$
\be\nonumber
\Psi:=\left(\frac{\gamma}{f}\right)^{1/12}={\sqrt\frac{a(t)}{a_0}}
\ee
and a background metric determinant $f$.
For the background space the present day sphere with the modern radius $a_0$ is taken:
\be\label{backgroundinterval}
f_{ij}{\rm d}x^i{\rm d}x^j=a_0^2\left({\rm d}\chi^2+\sin^2\chi ({\rm d}\theta^2+\sin^2\theta{\rm d}\phi^2)\right).
\ee
Hence, the conformal metric with components $\tilde\gamma_{ij}$ is the static metric of the modern Universe with radius $a_0$.
Under the conformal transformation (\ref{confmetric}) the Ricci scalar $R$ is transformed as
\be\label{R}
R=\Psi^{-4}\tilde{R}=\left(\frac{a_0}{a(t)}\right)^2\tilde{R}.
\ee
The conformal Hamiltonian constraint takes the form of the Lichnerowicz -- York equation
\be\label{LY}
-\frac{1}{9}K^2-\frac{1}{6}\Psi^{-4}\tilde{R}+\left(\frac{8\pi G}{3}\right)\rho=0.
\ee

The conformal Friedmann equation is obtained after the substitution in (\ref{LY})
the interval of the coordinate time by the interval of conformal time: $c{\rm d}t=a{\rm d}\eta$:
\begin{equation}\label{Friedmannconftime}
\left(\frac{{a}'}{a}\right)^2=\frac{8\pi G}{3}a^2\rho-{k},
\end{equation}
where the prime denotes the derivative with respect to the conformal time.
Since we will describe the physics in the conformal frame, we have to use the conformal time.
The conformal Friedmann equation becomes
\bea\label{Friedmannconformal}
&&\frac{1}{x^2}\left(\frac{{\rm d}x}{{\rm d}\eta}\right)^2=\\
&=&H_0^2\left[\Omega_{\rm \Lambda}x^2+\Omega_{\rm curv}+\frac{\Omega_{\rm M}}{x}+
\frac{\Omega_{\rm rad}}{x^2}+\frac{\Omega_{\rm rigid}}{x^4}\right].\nonumber
\eea
The energy continuity equation in conformal time takes the form:
\be\label{continuityconformal}
{\rho}'=-3(\rho+p)\left(\frac{{a}'}{a}\right).
\ee
The Hubble parameter in classical cosmology $H(t)$ and the conformal Hubble parameter ${\cal H}(\eta)$
are connected:
\be\nonumber
H(t)\equiv\frac{\dot{a}}{a}=\frac{a'}{a^2}\equiv\frac{1}{a}{\cal H}(\eta).
\ee

Leaving significant cosmological parameters for fitting the Hubble diagram, we get
\be\label{Friedmannconf}
\frac{1}{x^2}\left(\frac{{\rm d}x}{{\rm d}\eta}\right)^2=
\left(\frac{{\cal H}_0}{c}\right)^2\left[\frac{\Omega_{\rm M}}{x}+\frac{\Omega_{\rm rigid}}{x^4}\right],
\ee
where ${\cal H}_0=H_0$ is a conformal Hubble constant.
Interpretation of the Hubble diagram, based on conformal Friedmann equation with parameters
\be
\Omega_{\rm rigid} = 0.755,\qquad \Omega_{\rm M}=0.245,
\ee
yields the same qualitative approximation as the standard cosmological model with parameters
$\Omega_\Lambda = 0.72,$ $\Omega_{\rm M}=0.28$ ~\cite{Zakharov:2010nf}. A parameter $\Omega_{\rm rigid}$ corresponds to a rigid state matter, when the energy density is equal to the pressure, that happens under a nucleosynthesis regime in stars ~\cite{Narlikar}. This unusual equation of state arises when describing a quantum vacuum.
In quantum physics, the equation $p=\epsilon$ describes the Casimir vacuum of virtual massive boson and fermion particles ~\cite{EoS}. This confirms the manifestation of vacuum in the Hubble diagram.

If we identify the observed quantities with conformal quantities, then the evolution of lengths in cosmology is replaced by the evolution of masses ~\cite{Narlikar}.
The choice of the length scale is equivalent to the choice of the mass scale when describing physical processes in Nature.
The requirement of conformal symmetry leads to scaling of the masses of objects, which, in turn, leads to a new
(non-Doppler) interpretation of the cosmological redshift of atomic spectra.
The introduction of a relative standard means the identification of observable quantities with conformal fields and coordinates, which leads to a conformal mass cosmology instead of the standard distance evolution cosmology.
Photons emitted by atoms on distant stars billions of years ago ''remember`` the characteristics of the atoms. The wavelength of such a photon changes with the inverse electron mass, which is responsible for the transition.
Atoms were determined by their masses even at that distant time. Astronomers are currently comparing the spectrum with spectrum of the same atoms on Earth, but with an increased mass over the past time
$
m(\eta)=m_I a(\eta),
$
where $m_I$ is an initial mass of an atom, $a(\eta)$ is the conformal factor as a function of photon time $\eta$.
As a result, a redshift $z$ of the Fraunhofer spectral lines is observed
\be\nonumber
1+z=\frac{\lambda m_I}{[\lambda a(t)]m_I}=\frac{\lambda m_I}{\lambda [a(t)m_I]}.
\ee
The wavelength of the photon $\lambda$ does not change during all the time it moves in space, since its rest mass is equal to zero.
The cost of such an interpretation is negligible -- the masses of elementary particles turn out to be time-dependent, which is not comparable to the acceptance of the collapse of the space in which we live, so called Big Bang.

It is appropriate to remind the correct statement of Steven Weinberg
about interpretation of experimental data on redshifts ~\cite{WeinbergThree}:
\begin{quote}
``I do not want
to give the impression that everyone agrees with this interpretation of the red shift. We do not actually observe
galaxies rushing away from us; all we are sure of is that the lines in their spectra are shifted to the red, i.e.
towards longer wavelengths. There are eminent astronomers who doubt that the red shifts have anything to do with
Doppler shifts or with expansion of the universe''.
\end{quote}

From the conformal Friedmann equation (\ref{Friedmannconf}) with significant cosmological parameters  we obtain the formula for calculating the conformal time to a luminous object.
In conformal coordinates, the behavior of photons is exactly the same as in the Minkowski space. The time intervals between the emission of two photons and between their absorption are the same. The coordinate time interval
${\rm d}t=-a{\rm d}r$ and the conformal time interval ${\rm d}\eta = - {\rm d}r$ are different~\cite{GRub}.

The analytical formula for calculating the distance to a luminous object can be obtained after integrating (\ref{Friedmannconf}).
From the conformal Friedmann equation (\ref{Friedmannconf}) with significant cosmological parameters  we obtain the formula for calculating the conformal time to a luminous object
\be
\eta(z)=\frac{c}{H_0\sqrt{\Omega_{\rm M}}}\int\limits_{1/(1+z)}^1\frac{x{\rm d}x}{\sqrt{P(x)}},
\ee
where the polynomial $P(x)$ can be represented in a factorized form:
\be\nonumber
P(x)=x^3+q^3\equiv (x+q)(x^2-q x+q^2),
\ee
where $q\equiv\sqrt[3]{\Omega_{\rm rigid}/\Omega_{\rm M}}.$

The formula for calculating the distance to a luminous object includes the integral of the rational function
\be\label{CCLegandre}
I=\int\frac{x{\rm d}x}{\sqrt{P(x)}}.
\ee
Let us show how it can be expressed in terms of the elliptic integrals~\cite{Whittaker}.
We use the trigonometric substitution $x\mapsto\varphi$:
\be\nonumber
x=-q+\sqrt{3}q\tan^2\frac{\varphi}{2}.
\ee
After changing the variable, we get
\be\nonumber
P(\varphi)=\frac{3\sqrt{3}q^3\tan^2(\varphi /2)}{\cos^4(\varphi /2)}[1-k^2\sin^2\varphi],
\ee
where $k^2\equiv (2+\sqrt{3})/4.$
The differential is transformed according to the formula:
\be\nonumber
{\rm d}x=\sqrt{3}q\frac{\tan (\varphi /2)}{\cos^2(\varphi /2)}{\rm d}\varphi.
\ee
As a result of changing the variable, the integral (\ref{CCLegandre}) will take the form:
\bea\nonumber
I=&-&\sqrt{\frac{q}{\sqrt{3}}}\int\frac{{\rm d}\varphi}{\sqrt{1-k^2\sin^2\varphi}}\\
&+&\sqrt{\sqrt{3}q}
\int\frac{\tan^2(\varphi/2)\,{\rm d}\varphi}{\sqrt{1-k^2\sin^2\varphi}}.\label{intprCC}
\eea
We transform the second integral using the equality
\bea\nonumber
&&\int\frac{\tan^2(\varphi/2)\,{\rm d}\varphi}{\sqrt{1-k^2\sin^2\varphi}}=
2\tan(\varphi/2)\sqrt{1-k^2\sin^2\varphi}\\
&&-
2\int\sqrt{1-k^2\sin^2\varphi}\,{\rm d}\varphi +
\int\frac{{\rm d}\varphi}{\sqrt{1-k^2\sin^2\varphi}}.\nonumber
\eea
Then the integral (\ref{intprCC}) takes the form
\bea\nonumber
I&=&2\sqrt{\sqrt{3}q}\tan(\varphi/2)\sqrt{1-k^2\sin^2\varphi}\\
&-&2\sqrt{\sqrt{3}q}\int\sqrt{1-k^2\sin^2\varphi}\nonumber\\
&+&(\sqrt{3}-1)\sqrt{\frac{q}{\sqrt{3}}}
\int\frac{{\rm d}\varphi}{\sqrt{1-k^2\sin^2\varphi}}.
\nonumber
\eea
Substituting the limits of integration, we find the required distance
\bea
&&r(z)=\frac{2c}{H_0\sqrt{\Omega_{\rm M}}}\left[
\sqrt{(1+q)(1-k^2\sin^2\varphi (0))}\right.\nonumber\\
&&\left.-\sqrt{\frac{1+q (1+z)}{(1+z)}(1-k^2\sin^2\varphi (z))}\right.\nonumber\\
&&\left.+\frac{(\sqrt{3}-1)}{2}\sqrt{\frac{q}{\sqrt{3}}}\left[F(k,\varphi (0))-F(k,\varphi (z))\right]\right.\nonumber\\
&&\left.-\sqrt{\sqrt{3}q}\left[E(k,\varphi (0))-E(k,\varphi (z))\right]
\right].
\eea

Here, beside of the elliptic integral of the first kind $F(k,\varphi)$, the elliptic integral of the second kind according to Legendre $E(k,\varphi)$ is appeared
\be\nonumber
E(k,\varphi):=\int\limits_0^\varphi\sqrt{1-k^2\sin^2\phi}\,{\rm d}\phi,\qquad k^2\equiv\frac{2+\sqrt{3}}{4}
\ee
with modulus
$k^2$. The angle $\varphi$ is determined by the formula
\be\nonumber
\varphi (z)=2\arctan\sqrt{\frac{1+q (1+z)}{\sqrt{3}q (1+z)}}.
\ee

The conformal luminosity distance $d_L(z)_{CC}$ is related to the standard luminosity distance $d_L(z)_{SC}$ as~\cite{GRub}
\be\nonumber
d_L(z)_{CC}= (1+z)d_L(z)_{SC}=(1+z)^2r (z).
\ee
The effective magnitude -- redshift relationship in the conformal cosmology has the form
\begin{equation}\label{mMConf}
m(z)-M=5{\rm lg} [d_L(z)_{CC}]+{\cal M}.
\end{equation}

\section{Comparative analysis}

\begin{figure}[tbp]
\begin{center}
\includegraphics[width=3in]{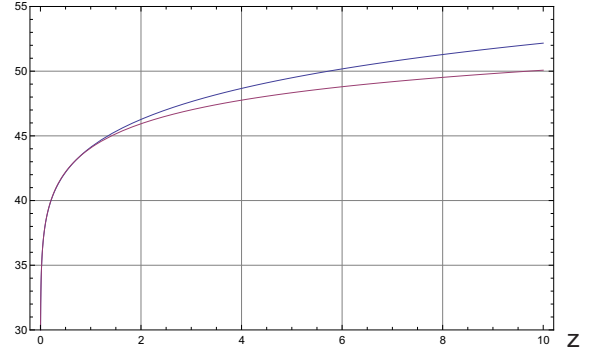}
\caption{\small Extrapolation of modern Hubble curves of two cosmological models: effective magnitude - redshift ratio. For large values of $ z $, the conformal model curve rises significantly above the standard model curve. The deviation exceeds the errors of astronomical observations. Future refined astronomical observations may suggest which model should be preferred.}
\label{ModernHubble}
\end{center}
\end{figure}

The analytical formulas (\ref{mMcl}), (\ref{mMConf}) are applicable to extrapolate the Hubble diagram for redshifts exceeding modern observational data on supernovae. For large values of $z$, the conformal model curve rises significantly above the standard model curve
(see Fig.~\ref{ModernHubble}). The deviation exceeds the errors of astronomical observations.
This discrepancy has already been noted in~\cite{Behnke}, where the problem been solved in the dilaton theory of gravity.

In the standard cosmology, there are characteristics of the Hubble diagram: Hubble parameter, a deceleration and a jerk~\cite{RiessNobel}
\begin{eqnarray}
H(t)&\equiv&+\left(\frac{\dot{a}}{a}\right)=H_0\sqrt{\frac{\Omega_{\rm M}}{a^3}+\Omega_\Lambda},\label{Hpar}\\
q(t)&\equiv&-\left(\frac{\ddot{a}}{a}\right)\left(\frac{\dot{a}}{a}\right)^{-2}=
\frac{\Omega_{\rm M}/2-\Omega_\Lambda a^3}{\Omega_{\rm M}+\Omega_\Lambda a^3},\label{qpar}\\
j(t)&\equiv&+\left(\frac{\dot{\ddot{a}}}{a}\right)\left(\frac{\dot{a}}{a}\right)^{-3}=1.\label{jpar}
\end{eqnarray}
The deceleration parameter $q$ changes its sign during the evolution of the Universe at the inflection point
\be\nonumber
a^{*}=\sqrt[3]{\frac{\Omega_{\rm M}}{2\Omega_\Lambda}},
\ee
$j$-parameter remains constant.
It should be noted that the value of the Hubble constant (\ref{Hubbleconstant}) varies quite a lot depending on the research methods: analysis of angular fluctuations of the microwave background or the use of traditional astronomical methods ~\cite{Dolgov}. But, the existence of the inflection point does not depend on the value $H_0$.

We define analogous parameters for the conformal time
\begin{eqnarray}
{\cal H}(\eta)&\equiv&+\left(\frac{{a}'}{a}\right),\label{Hparameter}\\
{q}(\eta)&\equiv&-\left(\frac{{a}''}{a}\right)\left(\frac{{a}'}{a}\right)^{-2},\label{qparameter}\\
{j}(\eta)&\equiv&+\left(\frac{{a}'''}{a}\right)\left(\frac{{a}'}{a}\right)^{-3}.\label{jparameter}
\end{eqnarray}
We calculate the conformal parameters using the conformal Friedmann equation~(\ref{Friedmannconf}).
Hubble parameter
\be\nonumber
{\cal H}(\eta)=\frac{{\cal H}_0}{a^2}\sqrt{\Omega_{\rm rigid}+\Omega_{\rm M} a^3}>0,
\ee
deceleration parameter
\be\nonumber
q(\eta)=\left(\frac{\Omega_{\rm rigid}-(\Omega_{\rm M}/2)a^3}{\Omega_{\rm rigid}+
\Omega_{\rm M}a^3}\right)>0,
\ee
hence, the scale factor grows with deceleration;
the jerk parameter
\be\nonumber
j(\eta)=\frac{3\Omega_{\rm rigid}}{\Omega_{\rm rigid}+\Omega_{\rm M}a^3}>0
\ee
changes from 3 to $3\Omega_{\rm rigid}$.
Dimensionless parameters $q(\eta)$ and $j(\eta)$ remain positive during the evolution. The Universe does not undergo the so-called jerk which is an artifact of approach of the standard cosmological model.
Future refined astronomical observations will suggest which model should be preferred.

\section{Discussion}

The conformal gravity theory has been advanced as a candidate to standard Einstein gravity~\cite{PhMannheim}.
There are hopes that an alternative to dark matter and dark energy is conformal gravity~\cite{Mann}.
In papers~\cite{tHooft1, tHooft2, tHooft3} a nontrivial connection between Einstein gravity and Weyl conformal gravity is found. 

\section*{Acknowledgement}

I am grateful to Professor A. B. Arbuzov for fruitful discussions on modern cosmological problems and the Laboratory of Theoretical Physics named after N.~N. Bogoliubov, Joint Institute for Nuclear Research (Dubna) for hospitality.
I obliged Professor J. David Brown for acquaintance with his paper on conformal invariance in General Relativity.
I express my gratitude for the interest in the work to the participants of ${\rm XVI}$-th International Conference ''Finsler Extension of Relativity Theory'', and of the seminar of the People's  Friendship University of Russia (Moscow).



\begin{thebibliography}{10}

\bibitem{RiessNobel}
A.~G.~Riess,
{Rev. Mod. Phys.} {\bf 84}, 1165 (2012).

\bibitem{Ram}
R.~G.~Vishwakarma, J.~V.~Narlikar,
Research in Astron. Astrophys. {\bf 10}, 1195 (2010).

\bibitem{Denzel}
Ph.~Denzel, et al., Monthly Notices of the Royal Astronomical Society. {\bf 501}, 784 (2020).

\bibitem{Soltis}
J.~ Soltis, et al., The Astrophysical Journal Letters. {\bf 908}, L5 (2021).

\bibitem{RiessApJL}
A.~G.~Riess, et al., The Astrophysical Journal Letters. {\bf 908}, L6 (2021).

\bibitem{Narlikar}
J.~V.~Narlikar, {\it Violent Phenomena in the Universe}
(Oxford University Press, 1984).

\bibitem{Zakharov:2010nf}
A.~F.~Zakharov, V.~N.~Pervushin,
{Int. J. Mod. Phys. D} {\bf 19}, 1875 (2010).

\bibitem{PavlovMIPh}
  A.~E.~Pavlov,
  {RUDN J. Math. Inform. Sc. Phys.} {\bf 25}, 390 (2017).

\bibitem{Fock}
V.~A.~ Fock, {\it Theory of Space, Time and Gravitation} (Pergamon Press, 1959).

\bibitem{ProblemsGauge}
M.~P.~D\c{a}browski, D.~Behnke, D.~ Blaschke.
In: {\it Problems of Gauge Theories: On occasion of 60th birthday of V.~N. ~Pervushin}. Eds.: B.~M.~ Barbashov, V.~V.~ Nesterenko. D2-2004-66 (JINR, Dubna, 2004).

\bibitem{Deser}
S.~ Deser, Annals of Physics, {\bf 59}, 248 (1970).

\bibitem{Feza}
F.~ G$\ddot{\rm u}$rsey, Annals of Physcis, {\bf 24}, 211 (1963).

\bibitem{Pavlov2017}
A.~E.~Pavlov, Grav. Cosmol. {\bf 23}, 208 (2017).

\bibitem{Brown}
J.~D.~ Brown, Phys. Rev. {\bf D 71}, 104011 (2005).

\bibitem{Whittaker}
E.~T.~Whittaker, G.~N.~Watson, {\it A Course of Modern Analysis} (Cambridge University Press, Cambridge, 1927).

\bibitem{EoS}
A.~E.~Pavlov, {Mod. Phys. Lett. A} {\bf 35}, 2050271 (2020).

\bibitem{WeinbergThree}
S.~ Weinberg, {\it The First Three Minutes: A Modern View of the Origin of the Universe} (Basic Books, 1977).

\bibitem{GRub}
D.~S.~Gorbunov, V.~A.~Rubakov, {\it Introduction to the Theory of the Early Universe: Hot Big Bang Theory}
(World Scientific, 2011).

\bibitem{Behnke}
D.~Behnke, et al., Phys. Lett. {\bf B 530}, 20 (2002).

\bibitem{Dolgov}
S.~I.~Blinnikov, A.~D.~Dolgov, {Physics -- Uspekhi} {\bf 62}, 529 (2019).

\bibitem{Shap}
M.~ Shaposhnikov, D.~Zenh$\ddot{\rm a}$usern,
Phys. Lett. {\bf B 671}, 187 (2009).

\bibitem{PhMannheim}
Ph.~ D.~ Mannheim, Prog. Part. Nucl. Phys. {\bf 56}, 340 (2006).

\bibitem{Mann}
Ph.~ D.~ Mannheim, Found. Phys. {\bf 42}, 388 (2012).

\bibitem{tHooft1}
G. 't Hooft, {\it Probing the small distance structure of canonical quantum gravity using the conformal group},
arXiv:1009.0669v2 [gr-qc].

\bibitem{tHooft2}
G. 't Hooft, {\it The conformal constraint in canonical quantum gravity},
arXiv:1011.0061v1 [gr-qc].

\bibitem{tHooft3}
G. 't Hooft, {\it A class of elementary particle models without any adjustable real parameters},
arXiv:1104.4543v1 [gr-qc].

\end{thebibliography}
\end{document}